\documentclass[structabstract]{aa}  
\pdfoutput=1 
\usepackage{natbib, graphicx}
\usepackage{amsmath}
\usepackage{subfigure}
\usepackage{lscape}
\usepackage{color}
\usepackage{txfonts}
\usepackage{multirow}
\usepackage{hyperref}
\begin{document}

   \title{Deriving precise parameters for cool solar-type stars}
   \subtitle{Optimizing the iron line list\thanks{Based on observations collected at the La Silla Paranal Observatory,
ESO (Chile) with the HARPS spectrograph at the 3.6-m telescope (ESO
runs ID 072.C-0488.}}
   \author{M. Tsantaki\inst{1,2} 
          \and    
          S. G. Sousa
          \inst{1,3}          
         \and    
          V. Zh. Adibekyan
          \inst{1}          
         \and    
          N. C. Santos 
         \inst{1,2}
         \and    
          A. Mortier 
          \inst{1,2} 
         \and 
          G. Israelian
          \inst{3}
           }
   \institute{Centro de Astrof\'{i}sica, Universidade do Porto, Rua das Estrelas, 4150-762 Porto, Portugal \\
              \email{Maria.Tsantaki@astro.up.pt} 
     \and
        Departamento de F\'{i}sica e Astronomia, Faculdade de Ci\^encias, Universidade do Porto, Rua do Campo Alegre, 4169-007 Porto, Portugal 
      \and
         Instituto de Astrof\'{i}sica de Canarias, E-38200 La Laguna, Tenerife, Spain 
          }

   \date{Received XXXX; accepted XXXX}
   \authorrunning{M. Tsantaki et al.}
   \titlerunning{Deriving precise parameters for cool solar-type stars}
   \abstract
   {Temperature, surface gravity, and metallicitity are basic stellar atmospheric parameters necessary to characterize a 
star. There are several methods to derive these parameters and a comparison of their results often shows considerable discrepancies, even 
in the restricted group of solar-type FGK dwarfs.}
    {We want to check the differences in temperature between the standard spectroscopic technique based on iron lines and the Infrared Flux 
Method (IRFM). We aim to improve the description of the spectroscopic temperatures especially for the cooler stars where the 
differences between the two methods are higher, as presented in previous work.}
    {Our spectroscopic analysis is based on the iron excitation and ionization balance, assuming Kurucz model atmospheres in LTE. The 
abundance analysis is determined using the code MOOG. We optimize the line list using a cool star (HD\,21749) with high resolution and 
high signal-to-noise spectrum, as a reference in order to check for weak, isolated lines.}
    {We test the quality of the new line list by re-deriving stellar parameters for 451 stars with high resolution and signal-to-noise 
HARPS spectra, that were analyzed in a previous work with a larger line list. The comparison in temperatures between this work and the 
latest IRFM for the stars in common shows that the differences for the cooler stars are significantly smaller and more homogeneously 
distributed than in previous studies for stars with temperatures below 5000\,K. Moreover, a comparison is presented between 
interferometric temperatures with our results that shows good agreement, even though the sample is small and the errors of the mean 
differences are large. We use the new line list to re-derive parameters for some of the cooler stars that host planets. Finally, 
we present the impact of the new temperatures on the [\ion{Cr}{i}/\ion{Cr}{ii}] and [\ion{Ti}{i}/\ion{Ti}{ii}] abundance ratios that 
previously showed systematic trends with temperature. We show that the slopes of these trends for the cooler stars become drastically 
smaller.}
   {}
  
\maketitle

\section{Introduction}
\label{intro}

Temperature ($T{}_{\mathrm{eff}}$), surface gravity ($\log g$), and metallicitity ($[Fe/H]$, where iron is used as a proxy) are basic 
atmospheric parameters necessary to characterize a star, as well as to determine other indirect and fundamental parameters, such as mass, 
radius and age that are acquired in combination with stellar evolutionary models \citep{girardi}. 
Precise and accurate stellar parameters are also essential in exoplanet searches. The light curve of a transiting planet orbiting a star 
gives information of the planetary radius always in dependence of the stellar radius (R$_{p}$ $\propto$ R$_{\star}$). Moreover, the mass 
of the planet, or the minimum mass in case the inclination of the orbit is not known, is determined from the radial velocity technique 
only if the mass of the star is known (M$_{p}$ $\propto$ M$_{\star}^{2/3}$). 
Therefore, the determination of the planetary radius and mass requires a combination of transiting and radial velocity data 
\citep{ammler, torres08, torres}. Apart from the derivation of planetary properties (mass, radius, density), stellar parameters can be used to reveal correlations between 
planets and their host stars that will give insights on their formation and evolution mechanisms. Several correlations have been studied so 
far with very interesting results, such as stellar metallicitity and planet frequency \citep[e.g.][]{san1, udry2, sousa3}, stellar 
metallicitity and planetary mass \citep{guillot}, stellar metallicitity and planetary orbital periods \citep{soz1}, stellar temperature 
and obliquities \citep{alb, winn}, metallicitity and planet radius \citep{buch}. It is obvious that in order to correctly characterize 
the planets and furthermore, to statistically verify such correlations, well-determined stellar parameters are required.

Hundreds of stellar spectra are available from radial velocity planet search programs \citep{udry1, vogt, mayor, bouchy, locurto}. These 
spectra are obtained with high resolution spectrographs and a combination of them gives high signal-to-noise (S/N) that makes the 
spectroscopic analysis more powerful.

The study of the H$\alpha$ wings \citep{fuhrmann}, the excitation and ionization balance of iron lines \citep{san1}, spectral synthesis 
techniques \citep{val}, line ratios (or line depth ratios) \citep{gray} are the basic spectroscopic techniques that along with 
fundamental techniques such as photometry and interferometry can be used for the determination of the effective temperature. A comparison 
between these different methods can show considerable discrepancies in their results. Even in the restricted group of solar-type stars, 
the effective temperatures obtained with these methods can differ significantly \citep[e.g.][]{kov1, ramirez, casa1, sousa1}. 

The temperature determination becomes more difficult when we focus on K-type stars. 
The difficulties in these stars with $T{}_{\mathrm{eff}}\lesssim$ 5000\,K emerge from their line crowed spectra that cause strong blending. 
Blending can be a considerable problem if one uses the standard technique based on the iron equivalent widths (EWs). Lines cannot be easily 
resolved and the continuum placement becomes more difficult, causing bad measurement of the EWs and hence, makes the calculation of stellar 
parameters ambiguous. Therefore, it is important to select carefully the iron lines in such manner 
that will eliminate the blending effects, especially for cool stars.
In addition, the choice of the atomic 
parameters influences the abundance determination. Some authors calculate the atomic parameters using the Sun as a reference to avoid the 
errors that emerge from the theoretical or laboratory values. In that way, the atomic parameters for stars that are different from the Sun, 
i.e. too hot or too cool can be no longer accurate enough.   

The accuracy in the fundamental parameters of planet host stars is of great importance to planetary studies since these errors can 
propagate to the planetary properties \citep[e.g.][]{pepe}. In addition, stellar parameters affect the determination of other element 
abundances ($[X/H]$) and can introduce potential biases in their abundance calculations \citep[e.g.][]{neves, adib}. 

In this paper, we use the data from previous work that performed the standard spectroscopic analysis for a sample of 
solar-type stars presented in the paper of \cite{sousa1} (hereafter SO08). This sample is part of the High Accuracy Radial velocity Planet 
Searcher (HARPS) Guaranteed Time Observations (GTO) survey that is composed of slow rotators and low activity FGK stars in order to detect 
low mass planets.

A comparison of these spectroscopic results with the Infrared Flux Method (IRFM) indicates a disagreement between the effective 
temperatures only for the cooler stars of the sample with temperatures below $\sim$\,5000\,K. Motivated by that, we compile an optimized line 
list to improve the accuracy of the stellar parameters for the cooler stars and compare our results with other independent methods (IRFM, 
interferometry). This paper is organized as follows: in Sect.~\ref{sample1} we describe the stellar sample, the spectroscopic method and the criteria 
to select an optimized line list. We compare our results with the previous work of SO08 in Sect.~\ref{new}. In Sect.~\ref{ion}, we compare 
the spectroscopic surface gravities with the trigonometric ones and address the ionization problem for the cooler stars of our sample. In 
Sect.~\ref{other}, we compare the new derived parameters with the IRFM and interferometry. In Sect.~\ref{planets}, we calculate new 
parameters for some cool planet hosts. Finally, in Sect.~\ref{elements}, we use the new parameters for the cooler stars to reduce the trends of 
other elements ([\ion{Cr}{i}/\ion{Cr}{ii}] and [\ion{Ti}{i}/\ion{Ti}{ii}] abundance ratios) with temperature, as reported in \cite{adib}.

\section{Stellar sample and previous spectroscopic analysis}\label{sample1}

The stellar sample, presented in SO08, is composed of 451 stars as part of the HARPS high-precision 
GTO program at the ESO La Silla 3.6m telescope with the objective to 
detect low-mass extra-solar planets with high radial velocity accuracy \citep{mayor}. 
It is mainly comprised of dwarf FGK stars selected from a volume-limited
sample of the CORALIE survey \citep{udry1}. Planet hosts from the southern 
hemisphere were also added to this sample, forming in total a sample of 451 stars.   
These stars are slowly-rotating, non-evolved, and low-activity stars, with
apparent magnitudes that range from 3.5 to 10.2 and have distances of less 
than 56 parsec. The spectra have a resolution of R\,$\sim$\,110,000 and 90\% of the combined spectra 
have S/N higher than 200. We point to SO08 for more details.

  \begin{table}
     \centering
     \caption[]{Characteristics of the sample and the reference star, described in SO08.}
     \label{sample}
    $$ 
        \begin{array}{p{0.3\linewidth}ccccc}
           \hline\hline
           & T{}_{\mathrm{eff}} & \log g & [Fe/H] & Mass \\
           & (K) & (dex) & (dex) & M_{\odot} \\
           
           \hline
           Lowest & 4556  & 3.68  & -0.84 & 0.37 \\ 
           Highest & 6403 & 4.62 & 0.39 & 1.42  \\
           Reference star & 4723 & 4.40 & -0.02 & 0.76 \\
           \hline
        \end{array}
    $$ 
  \end{table}
 
For this sample, SO08 derived stellar parameters by imposing excitation and ionization equilibrium, based on the measurements of weak iron 
lines. This method is very effective for FGK stars due to the numerous iron lines in their spectra. Iron abundance is used as a proxy for 
the overall stellar metallicitity. 

The line list for their spectroscopic analysis, was composed of 263 \ion{Fe}{i} and 36 \ion{Fe}{ii} lines. The EWs of the lines were 
measured automatically for all stars using the ARES\footnote[1]{The ARES code is an open source code and can be found at 
http://www.astro.up.pt/$\sim$sousasag/ares} code \citep[Automatic Routine for line Equivalent widths in stellar Spectra;][]{sousa2}. 
The atomic parameters of the iron lines, namely the oscillator strength values ($\log gf$), were computed by an inverted solar analysis, 
using a solar model with $T{}_{\mathrm{eff}}$ = 5777\,K, $\log g$ = 4.44 dex, $\xi_{t}$= 1.0 m $s^{-1}$, $\log _{\epsilon}(Fe)$ = 
7.47 dex. 

The spectroscopic analysis was completed assuming Local Thermodynamic Equilibrium (LTE), and using the 2002 version of the 
abundance determination code MOOG\footnote[2]{The MOOG code can be downloaded free at http://verdi.as.utexas.edu/moog.html} \citep{sneden} 
and a grid of Kurucz Atlas 9 plane-parallel, 1D static model atmospheres \citep{kurucz}. 

The best parameters were obtained when the \ion{Fe}{i} abundance shows no dependence on the 
excitation potential, $\chi_{l}$, and on the reduced equivalent width, $\log W_{\lambda}/\lambda$.
Additionally, the mean abundances given by \ion{Fe}{i} and the \ion{Fe}{ii} must be the same (ionization balance) and consistent with those 
of the input model atmosphere. The input parameters converge to the true ones with an iterative minimization code based on the Downhill 
Simplex Method \citep{press}, making the total procedure automatic \citep{san1}. Some characteristics of the sample are depicted in 
Table~\ref{sample}, as described in SO08.

There are different sources of uncertainties that occur in the stellar parameter determination using this method. 
These errors can be attributed to the uncertainties of the measurements of the EWs, the uncertainties in the atomic parameters and 
the uncertainties that are intrinsic to the method of ionization and excitation equilibrium. 
In addition, systematic errors can arise due to the assumptions of the method, such as 1D static atmospheres, NLTE effects 
\citep{mashonkina,bergemann}. However, departures from LTE for Fe lines do not need to be considered for near solar metallicity 
dwarfs but should be taken into consideration for more evolved or very metal-poor stars \citep{lind,ruchti} that are not part of this 
sample. 

Errors in the measurements of the EWs can be minimized by using high quality spectra. In low S/N spectra, weak lines cannot be distinguished 
from noise and strong lines can be underestimated due to the miscalculation of their wings. The high resolution and high S/N spectra used for 
this sample, are the best solution to deal with such errors. Since in our spectroscopic analysis the atomic data ($\log gf$) are derived 
with respect to the Sun, we expect small errors for solar analogs and more significant for cooler and hotter stars. 

The errors in the stellar parameters for this analysis are calculated by varying each parameter (temperature, surface gravity and 
microturbulence) by a standard value. The uncertainty in $\xi_{t}$ is determined from the standard deviation in the slope of the 
least-squares fit of $\log_{\epsilon}(\ion{Fe}{i})$ versus $\log W_{\lambda}/\lambda$. The uncertainty in $T{}_{\mathrm{eff}}$ is 
determined from the uncertainty in the slope of the least-squares fit of $\log_{\epsilon}(\ion{Fe}{i})$ versus $\chi_{l}$ in addition to 
the uncertainty in the slope due to the uncertainty in $\xi_{t}$. The uncertainty in $\log g$ is derived from the contribution of the 
uncertainty in $T{}_{\mathrm{eff}}$ and microturbulence in addition to the scatter error of the \ion{Fe}{ii} abundance (measured as 
$\sigma$/$\sqrt{N}$, $\sigma$ is the standard deviation and N the number of lines). 
The uncertainty in the 
$\log_{\epsilon}(\ion{Fe}{i})$ is the sum of the squared uncertainties due to the error in $T{}_{\mathrm{eff}}$, $\xi_{t}$, and the 
scatter error of \ion{Fe}{i} abundance. The use of many iron lines can reduce this type of uncertainty, assuming 
that the majority of the lines are independent and of good quality.

\subsection{Building a stable line list for the cooler stars}\label{build}

A reliable line list is comprised of lines that can be accurately measured, which usually means unblended lines. In addition, 
lines must be unsaturated, cover a wide range in excitation potential and have accurate atomic data. 
Temperature, as well as the other stellar parameters, is strongly correlated with the equivalent width. 
This sensitivity emerges from the excitation and ionization processes that follow the exponential and power dependencies with temperature 
that are defined by the well-known Boltzmann and Saha equations. 

For the cooler stars line blending is severe, which makes the measurements of the EWs problematic. In particular, blending 
effects cause an overestimation of the EWs as two blended lines cannot be resolved. 

  \begin{figure}
   \centering
   \includegraphics[width=1.0\linewidth]{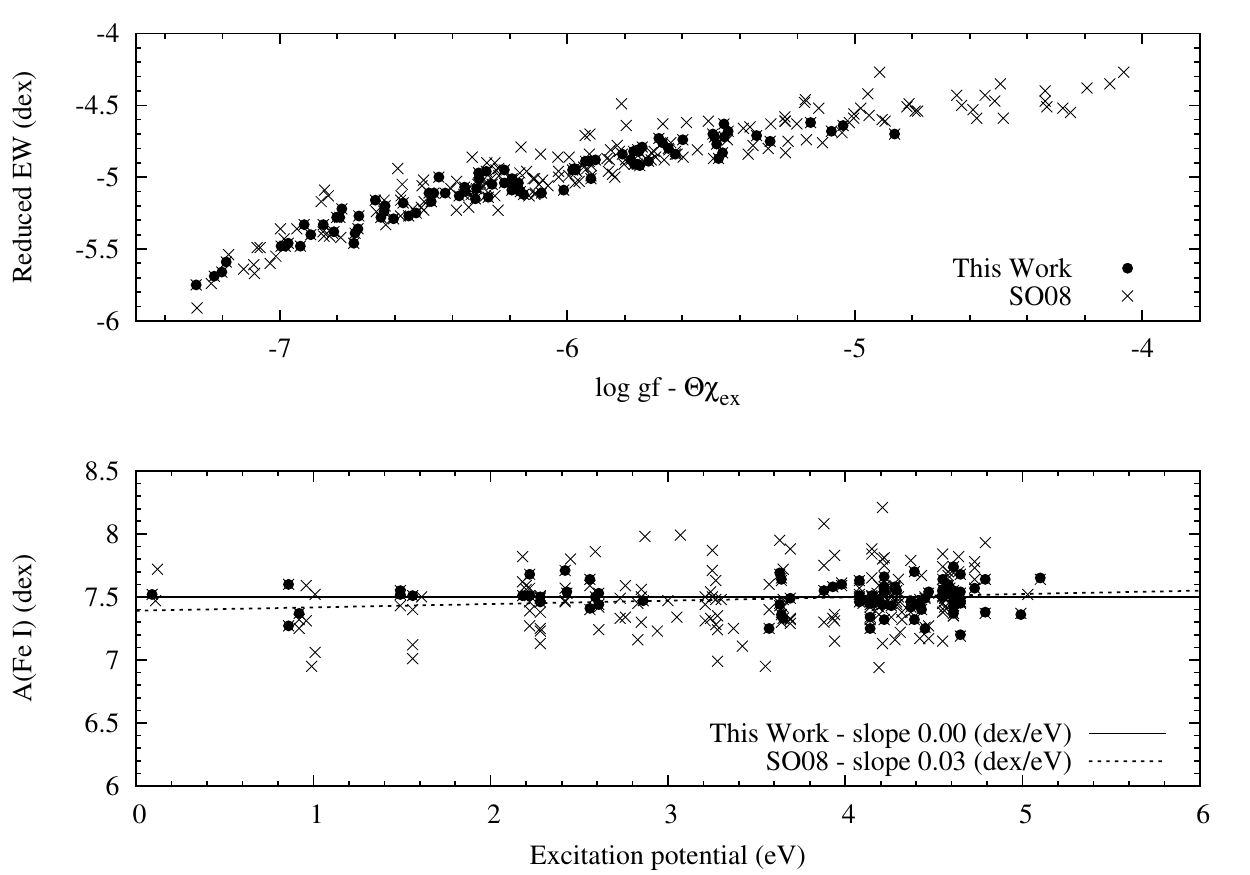}
   \caption{\footnotesize{ Upper panel: Curve of growth for both line lists for the reference star computed for the temperature 
($\Theta$=5040/$T{}_{\mathrm{eff}}$) of this work. Circles represent the reduced EW of the line list of this work and crosses 
the line list of SO08. Lower panel: The \ion{Fe}{i} abundances of the reference star versus the excitation potential. The dashed 
line shows the positive slope that corresponds to the line list of SO08. The solid line corresponds to the line list of this work and the 
slope is obviously zero. }}
   \label{line}
   \end{figure}

Another bias in the EW measurements may come from the fitting of strong lines. Gaussian fitting is a good approximation for weak lines 
and it can be reliable up to 200m\AA{} based on our experience, whereas a Voigt profile should be used for stronger lines.
Saturated lines that deviate significantly from the linear part of the curve of growth should also be avoided in the abundance analysis.
The EW predicted by the models of strong lines that are highly saturated, is quite dependent on microturbulence. A wrong estimation 
of microturbulence, will then produce errors in the abundance of any highly saturated line. 
 
On the other hand, weak lines that are strongly blended could lead to a sub-estimation of the continuum and consequently of the EW. This effect, 
however, is less significant. An overestimation in the EWs due to blending, as well as the underestimation of very strong lines could be 
the reason for the systematic raise in temperature that is observed for the cooler stars of SO08.
In addition, the reduced equivalent width could also be affected by such biases, leading to correlations with the excitation 
potential (see Appendix~\ref{app_re}).

  \begin{table}
     \centering
     \caption[]{Sample of the line list used for the spectroscopic analysis with the atomic parameters of \ion{Fe}{i} and \ion{Fe}{ii} as 
well as the corresponding EWs of the reference star HD\,21749.}
     \label{linelist}
    $$ 
        \begin{array}{p{0.2\linewidth}ccccc}
           \hline\hline
           $\lambda$ (\mbox{\AA}) & \chi_{l} & \log gf & Element & EW (m\mbox{\AA}) \\
           \hline
4508.28 & 2.86 & -2.403 & \ion{Fe}{ii} & 53.0 \\
4520.22 & 2.81 & -2.563 & \ion{Fe}{ii} & 72.4 \\
4523.40 & 3.65 & -1.871 & \ion{Fe}{i} & 101.7 \\
4537.67 & 3.27 & -2.870 & \ion{Fe}{i} & 43.2 \\
4551.65 & 3.94 & -1.928 & \ion{Fe}{i} & 41.9 \\
4556.93 & 3.25 & -2.644 & \ion{Fe}{i} & 57.9 \\
4566.52 & 3.30 & -2.156 & \ion{Fe}{i} & 68.6 \\
4574.22 & 3.21 & -2.353 & \ion{Fe}{i} & 55.1 \\
4576.34 & 2.84 & -2.947 & \ion{Fe}{ii} & 29.6 \\
 ... & ... & ... & ... & ... \\ 
           \hline
        \end{array}
    $$ 
  \end{table}

Therefore, our aim is to optimize the iron line list of SO08. With this goal, we use the K-type dwarf HD\,21749 with $T{}_{\mathrm{eff}}$ = 
4723\,K (see Table~\ref{sample}), as reference in order to check for unblended lines in its high S/N spectrum. After visual inspection, 
we only consider weak, isolated lines that can give good estimation for the local continuum. We avoid strong lines ($>$200 m\AA{}) in 
order to apply Gaussian profiles. For the reference star we show the curve of growth \citep[see][]{gray} using both line lists 
(Fig.~\ref{line} upper panel). Limiting the EW cut off, we mitigate in large amount the problem of saturated lines and microturbulence. 
The proof of that mitigation is the fact that the derived temperatures with the new line list agree with other less model dependent methods 
(see Sect.~\ref{other}).

Very weak lines ($<$10 m\AA{}) were also excluded so that noise is not superposed to these lines. The region of the 
spectrum below 4500 \AA{} is neglected due to the higher blending. The final line list\footnote[3]{The full line list is available in 
electronic form.} is compiled with 120 \ion{Fe}{i} and 17 \ion{Fe}{ii} lines, as shown in Table~\ref{linelist}. 

As mentioned before, the effective temperature is derived when the correlation coefficient between $\log_{\epsilon}(\ion{Fe}{i})$ and 
$\chi_{l}$ is zero. In the lower panel of Fig.~\ref{line}, we demonstrate this correlation for the reference star using the line list of 
this work and of the work of SO08 using the parameters derived with the line list of this work. The positive slope for the line list of 
SO08 is translated in an overestimation in temperature of $\sim$180 K for this star. In addition, the abundances with the new line list 
show a smaller scatter which corresponds to smaller errors in the final temperature value.

\section{New stellar parameters for 451 FGK stars in the HARPS GTO sample}\label{new}

\begin{table*}
\centering
\caption{Part of the derived stellar parameters of the 451 stars.}
\begin{tabular}{lcccccc}
\hline\hline
Star & $T{}_{\mathrm{eff}}$ & $\log g$  & $\xi_{t}$ & $[Fe/H]$ & Mass & Age \\
     & (K) & (cm s$^{-2}$) & (km s$^{-1}$) &  (dex) & (M$_{\odot}$) & (Gyr) \\
\hline
... & ... & ... & ... & ... & ... & ... \\
HD 967 & 5595$\pm$17 & 4.59$\pm$0.02 & 0.90$\pm$0.05 & -0.66$\pm$0.01 & 0.76$\pm$0.02 & 7.8$\pm$3.9 \\
HD 1237 & 5489$\pm$39 & 4.46$\pm$0.11 & 1.04$\pm$0.06 & 0.06$\pm$0.03 & 0.92$\pm$0.02 & 1.8$\pm$1.7 \\
HD 1320 & 5699$\pm$12 &4.55$\pm$0.05 & 0.89$\pm$0.02 & -0.26$\pm$0.01 & 0.91$\pm$0.02 & 2.4$\pm$2.0 \\
HD 1388 & 5970$\pm$15 & 4.42$\pm$0.05 & 1.13$\pm$0.02 & 0.00$\pm$0.01 & 1.06$\pm$0.01 & 3.0$\pm$0.8 \\
HD 1461 & 5740$\pm$16 & 4.36$\pm$0.03 & 0.94$\pm$0.02 & 0.18$\pm$0.01 & 1.05$\pm$0.02 & 2.0$\pm$1.1 \\
HD 1581 & 5990$\pm$15 & 4.49$\pm$0.07 & 1.24$\pm$0.03 & -0.18$\pm$0.01 & 1.03$\pm$0.02 &2.2$\pm$1.0 \\
HD 2025 & 4851$\pm$49 & 4.49$\pm$0.13 & 0.51$\pm$0.18 & -0.37$\pm$0.02 & 0.71$\pm$0.01 & 4.6$\pm$4.0 \\
HD 2071 & 5729$\pm$12 & 4.49$\pm$0.02 & 0.93$\pm$0.02 & -0.08$\pm$0.01 & 0.99$\pm$0.01 & 0.9$\pm$0.6 \\
HD 2638 & 5169$\pm$77 & 4.41$\pm$0.15 & 0.66$\pm$0.13 & 0.12$\pm$0.05 & 0.85$\pm$0.03 & 4.0$\pm$3.8 \\
... & ... & ... & ... & ... & ... & ...\\
\hline
\label{myresults}
\end{tabular}
\end{table*}

In order to check the effectiveness of the new line list, we re-derive stellar parameters for the 451 stars of the sample. For consistency, 
we use the same EWs as in SO08 that were measured automatically for all stars with the ARES code. In addition, we use the same damping 
parameters and atomic data. After a preliminary determination of the fundamental parameters, we perform a '3$\sigma$ clipping' procedure for 
lines that contribute with abundances higher than 3$\sigma$ from the average abundance. This procedure was also applied in SO08.

Table~\ref{myresults} shows the final parameters for a fraction of stars from the sample. Microturbulence ($\xi_{t}$) is used as a free 
parameter and is also derived from this spectroscopic analysis. The correlation of microturbulence with temperature and surface gravity 
is presented in Appendix~\ref{app}. This calibration can be useful in cases where the value of $\xi_{t}$ is set fixed. The stellar masses are 
calculated using the stellar evolutionary models from the Padova group\footnote[4]{Web interface for stellar mass estimation: 
http://stev.oapd.inaf.it/cgi-bin/param}. The errors of the fundamental parameters are internal, attributed to the method. They, thus, 
represent relative errors and not the absolute accuracy.  

\subsection{Internal comparison}

 \begin{figure}
  \centering
   \includegraphics[width=1.0\linewidth]{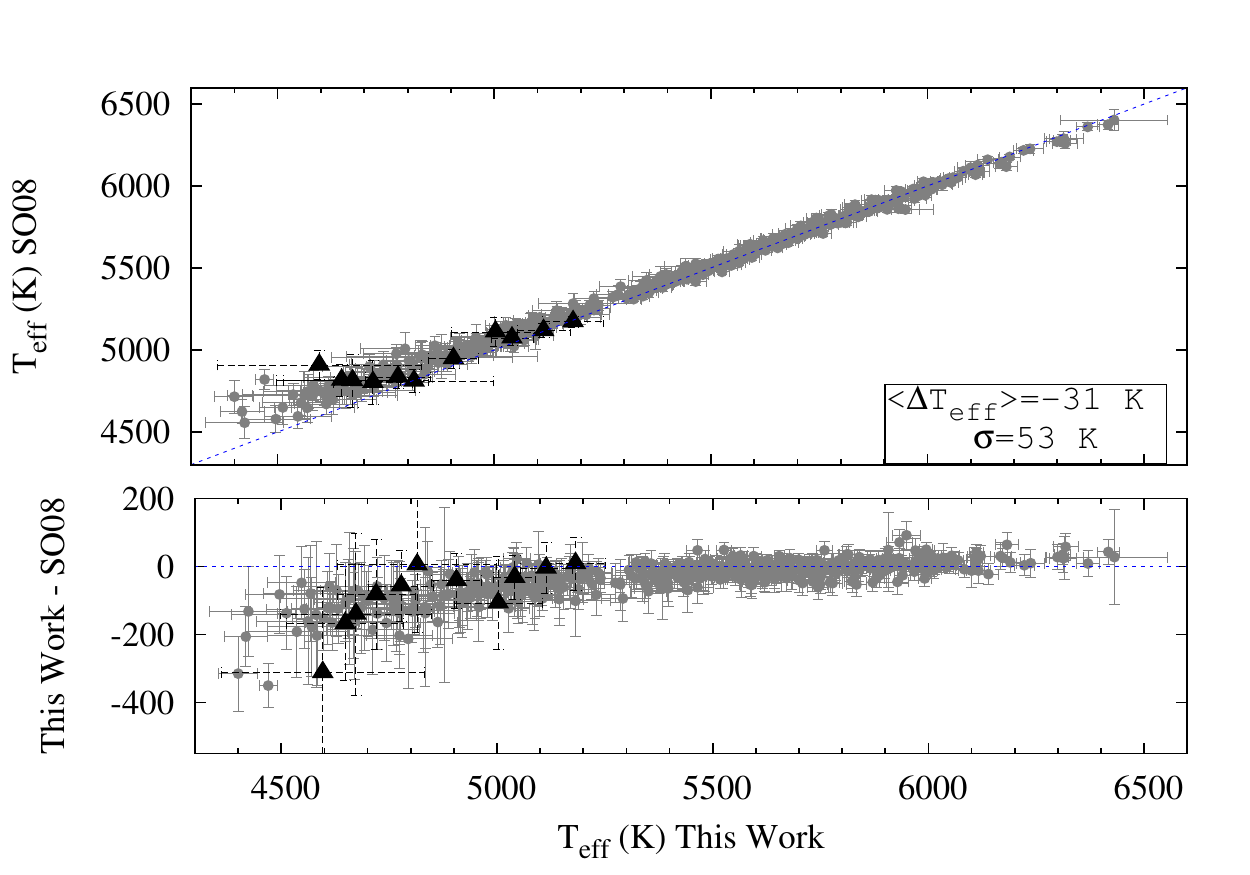}
   \includegraphics[width=1.0\linewidth]{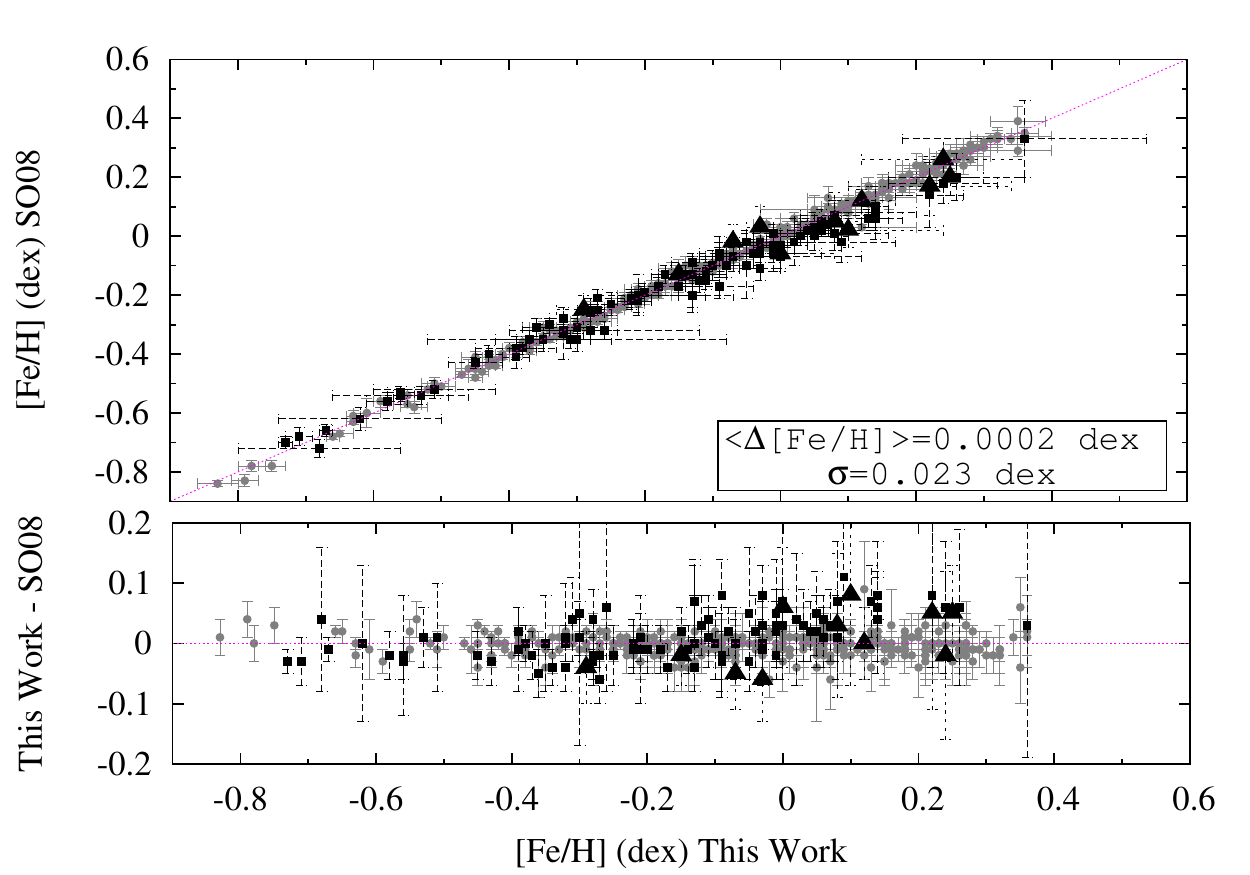}
   \includegraphics[width=1.0\linewidth]{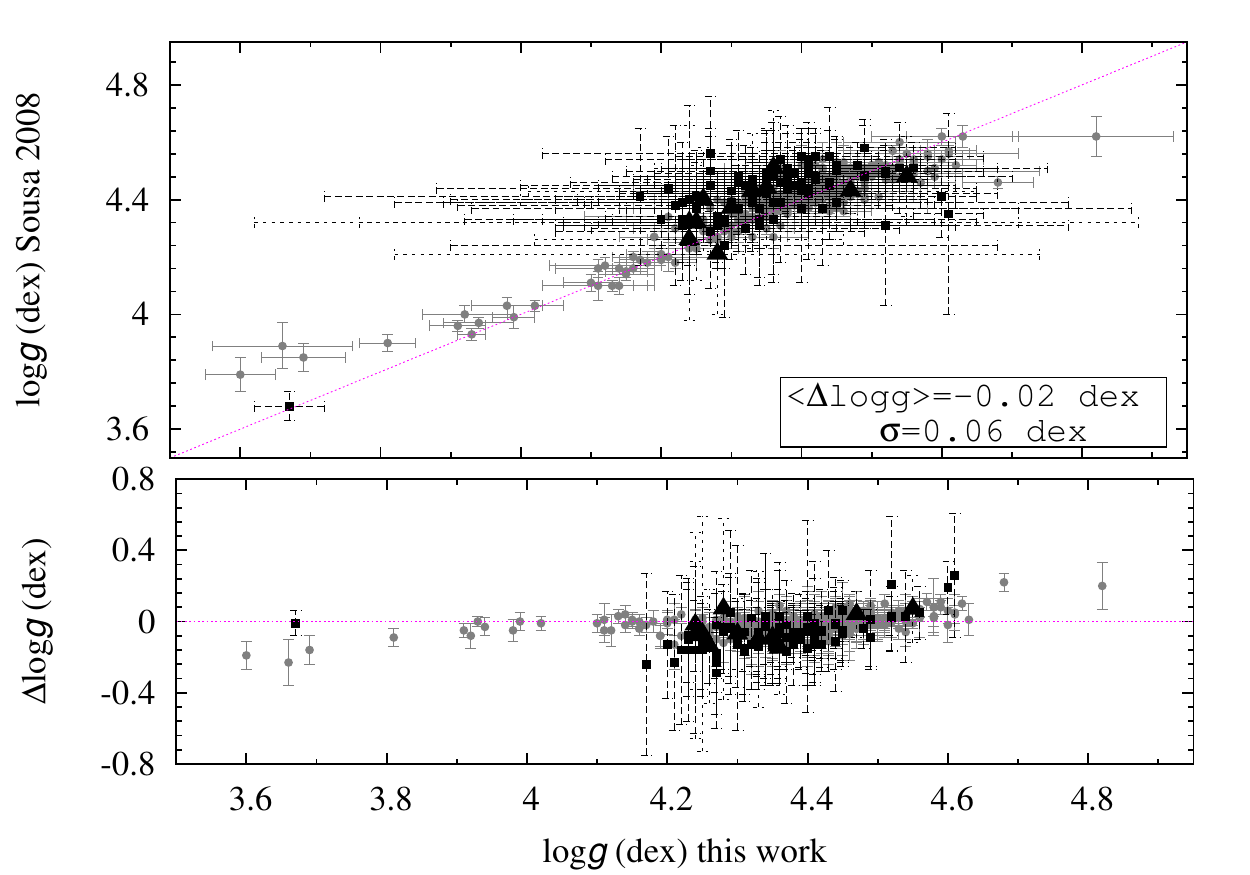}
  \caption{\footnotesize{Comparison between the parameters derived with the cool line list of this work and the results of SO08 for: 
temperature (top panel), metallicitity (middle panel) and surface gravity (bottom panel). $\Delta T{}_{\mathrm{eff}}$ corresponds to this 
work minus SO08. Black squares represent stars with $T{}_{\mathrm{eff}}$$<$5000\,K. Triangles represent stars with planets taken from 
Table~\ref{TabHosts} (see Sect.~\ref{planets}).}}
  \label{fig:1}
  \end{figure}

   \begin{figure*}
   \centering
\mbox{\subfigure{\includegraphics[width=3.1in]{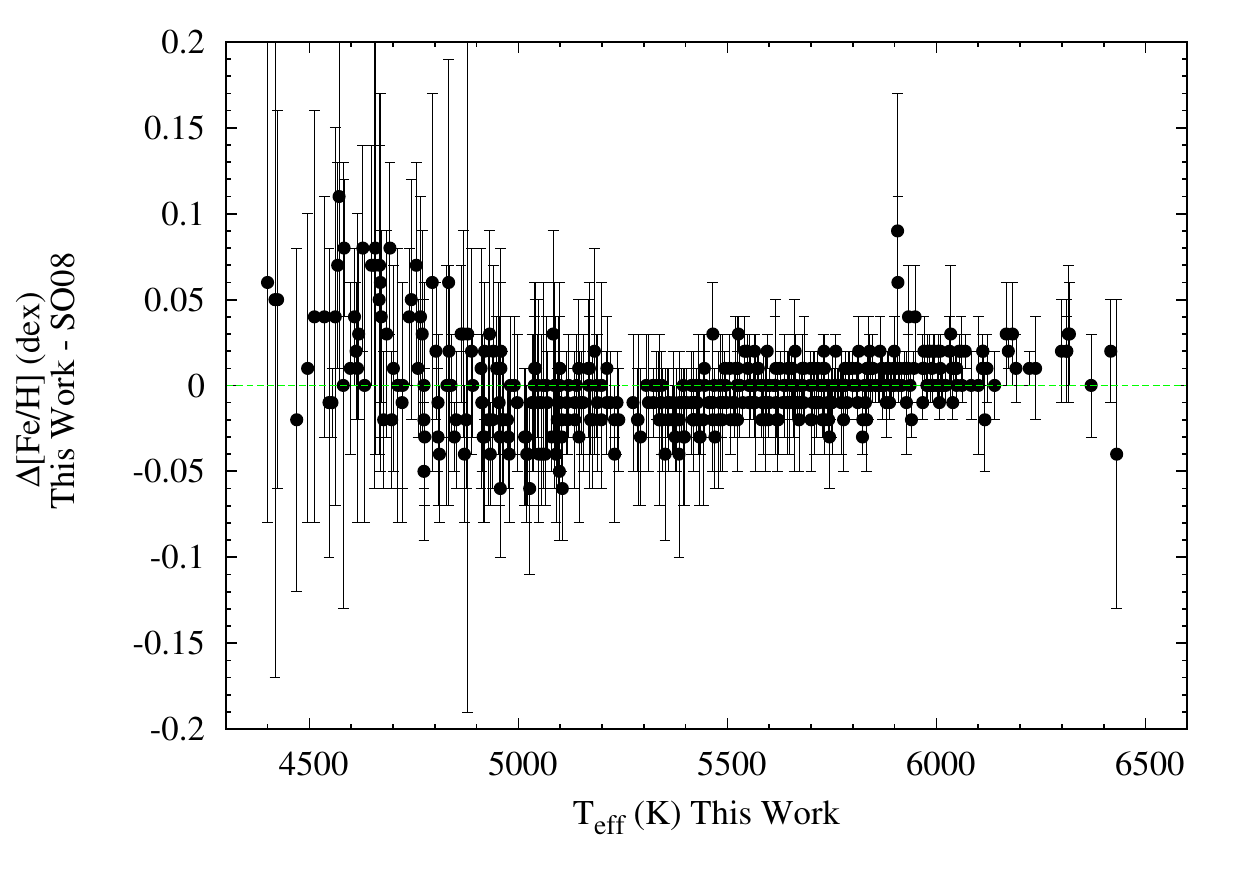}}\quad
\subfigure{\includegraphics[width=3.1in]{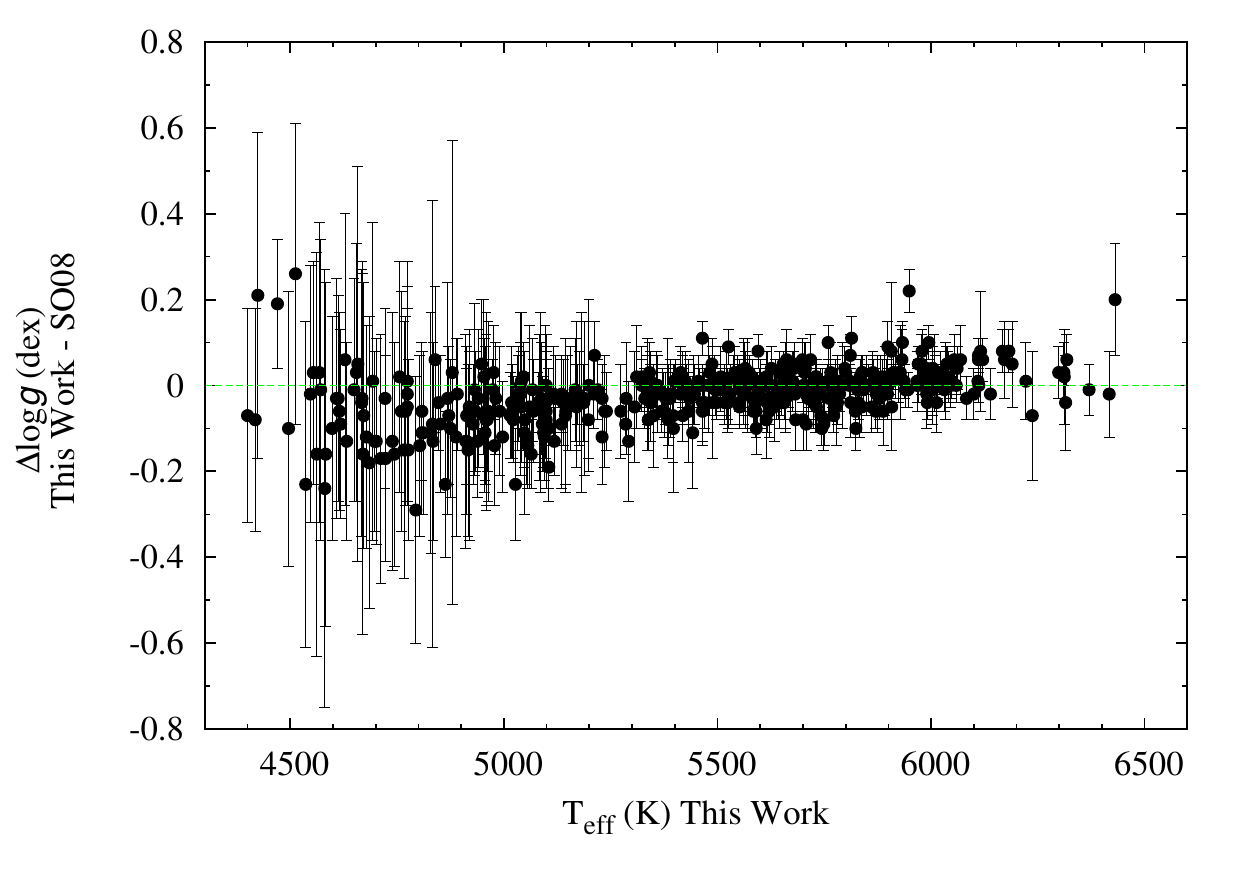} }}
\caption{\footnotesize{The effect of temperature on the other parameters: metallicitity (left panel) and surface gravity 
(right panel). The differences correspond to the values of this work minus the results of SO08.
}} 
   \label{fig:2}
   \end{figure*}

  \begin{table}
     \centering
     \caption[]{Results of the internal comparison for the whole sample and for stars with $T{}_{\mathrm{eff}}$$<$5000\,K.}
     \label{internal_comparison}

        \begin{tabular}{p{0.30\linewidth}cccc}
           \hline\hline
           & $\Delta T{}_{\mathrm{eff}}$ & $\Delta \log g$ & $\Delta$ [Fe/H]  \\
           & (K) & (dex) & (dex) \\
\hline
This Work -- SO08 & \multirow{2}{*}{-31$\pm$3} & \multirow{2}{*}{-0.023$\pm$0.003} & \multirow{2}{*}{0.0002$\pm$0.0011} \\
   ~~ whole sample    &   &  &        \\  
This Work -- SO08 & \multirow{2}{*}{-106$\pm$6} & \multirow{2}{*}{-0.07$\pm$0.01} & \multirow{2}{*}{0.0127$\pm$0.0001} \\ 
  ~~ $T{}_{\mathrm{eff}}<$5000\,K &         &              &              \\  
         \hline
         \hline
        \end{tabular}

  \end{table}
 
We present the comparison results with the work of SO08 for temperature, surface gravity and metallicitity, respectively in Fig.~\ref{fig:1}. 
Temperatures show very good agreement in the high and intermediate ranges. Table~\ref{internal_comparison} shows the mean difference in 
$T{}_{\mathrm{eff}}$ for the whole sample and for temperatures below 5000\,K with their standard errors\footnote[5]{The standard errors of 
the mean ($\sigma_{M}$) are calculated with the following formula: $\sigma_{M}$=$\frac{\sigma}{\sqrt{N}}$, $\sigma$ being the standard 
deviation.}. The mean difference in $T{}_{\mathrm{eff}}$ is -31$\pm$3 ($\sigma$=53)\,K for the whole temperature range. The significant 
differences appear, as expected, for stars with temperatures below 5000\,K with mean difference $\Delta T{}_{\mathrm{eff}}$ = -106$\pm$6 
($\sigma$=54)\,K. 

One interesting result is that even though we have considerably large differences in the low temperature regime, the values of 
metallicitity remain unaffected with $<\Delta [Fe/H]>$=0.00$\pm$0.00 ($\sigma$=0.02) dex for the overall sample. The same effect appears 
for surface gravity with $<\Delta \log g>$=-0.02$\pm$0.00 ($\sigma$=0.06) dex, even though there is bigger scatter. This result suggests 
that surface gravity and metallicity are not as sensitive to the selection of the line list as temperature, for this temperature regime and 
for this method. The same effect appears for the cool stars as seen in Table~\ref{internal_comparison}.

The impact of the updated effective temperatures on $[Fe/H]$ and on $\log g$ is depicted on Fig.~\ref{fig:2}. The changes in metallicitity 
show almost no correlation, within the errors, with the effective temperature. Only slightly higher metallicities appear for low temperatures, 
when comparing with SO08, yet within the errors. The surface gravities are also not correlated with temperature, even though there is high 
dispersion in the low temperature region. This result suggests that using this technique, a potential error in one of the parameters will 
not propagate to the others, avoiding systematic errors \citep[see also][]{torres}.

Metallicity has a key role in planet formation theories and is correlated with the planet frequency. The stellar sample, as 
mentioned before, is part of the HARPS GTO planet search program and it contains 102 up-to-date planet hosts. The metallicitity distribution of the 
sample, presented in SO08, shows that the Jovian planets are preferentially found in metal-rich stars, in contrast to Neptune-like planets 
that do not seem to follow this trend, even though the number of these planets is small. The new metallicities derived with the new line 
list do not change this trend, making this correlation between stars and planets reliable even before adapting the new temperatures.

\section{Ionization balance problem in cool stars.}\label{ion}

In an LTE abundance analysis for solar type stars the ionization equilibrium should be satisfied. Using the standard spectroscopic 
method we force the abundance of \ion{Fe}{i} and \ion{Fe}{ii} to agree. Surface gravity is determined from this tuning. However, there are 
many studies for cluster stars \citep{yong, morel, schuler} and field stars \citep{allende, ramirez07, ramirez13} showing that the cooler 
dwarf stars deviate from the ionization balance, with systematic higher \ion{Fe}{ii} abundances over \ion{Fe}{i}. The authors explain these 
discrepancies due to possible different scales in the stellar parameters (namely $T{}_{\mathrm{eff}}$ and $\log g$) 
or due to NLTE effects caused by the simplifications of the model atmospheres. The ionization balance of \ion{Fe}{i} and \ion{Fe}{ii} can
be investigated in this work by the behaviour of surface gravity. In fact, surface gravity mostly depends on \ion{Fe}{ii} lines, once 
the temperature scale is correct. 
 
An essential test for the accuracy of surface gravity derived from spectroscopy is the comparison with surface gravity derived from 
parallaxes (trigonometric $\log g$), based on the fundamental relation:
 
 \begin{equation}
  \log \frac{g}{g_{\odot}} = \log \frac{M}{M_{\odot}} + 4 \log \frac{T_{eff}}{T_{eff, \odot}} - \log \frac{L}{L_{\odot}} \nonumber
 \end{equation}
  
We calculated the trigonometric $\log g$ using the new \textit{Hipparcos} parallaxes \citep{hip}, V magnitudes, bolometric correction 
based on \cite{flower} and \cite{torress}, solar magnitudes from \citep{bessell}, the spectroscopic masses and $T{}_{\mathrm{eff}}$. No 
correction for interstellar reddening is needed since all stars are less than 56 pc in distance. 

 \begin{figure}
   \centering
   \includegraphics[width=1.0\linewidth]{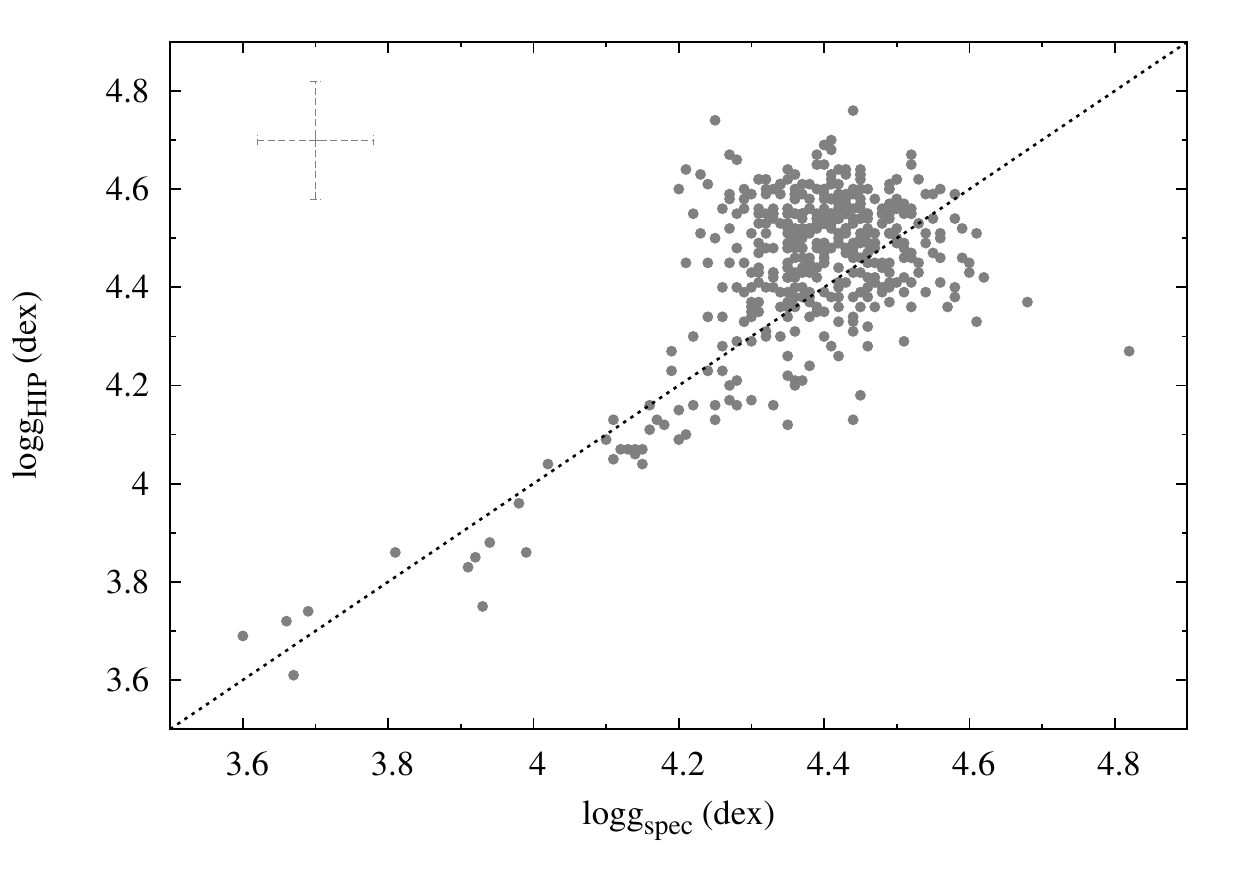}
  \caption{\footnotesize {Comparison of the surface gravities derived from spectroscopy, $\log g_{spec}$ of this work and from 
\textit{Hipparcos} parallaxes, $\log g_{HIP}$. A mean error for both axes is given in the upper left part.}}
  \label{logg_hip}
  \end{figure}

In Fig.~\ref{logg_hip}, we compare the spectroscopic $\log g_{spec}$ with the trigonometric $\log g_{HIP}$. At first glance, the 
spectroscopic $\log g$ agrees with the trigonometric ($<\log g_{HIP} - \log g_{spec}>$=0.07 dex). However, a more careful look shows that 
there is a disagreement especially for the high values of $\log g$ ($>$4.5 dex) where the spectroscopic gravities are underestimated, 
which is also observed in the work of SO08.

 \begin{figure}
   \centering
   \includegraphics[width=1.0\linewidth]{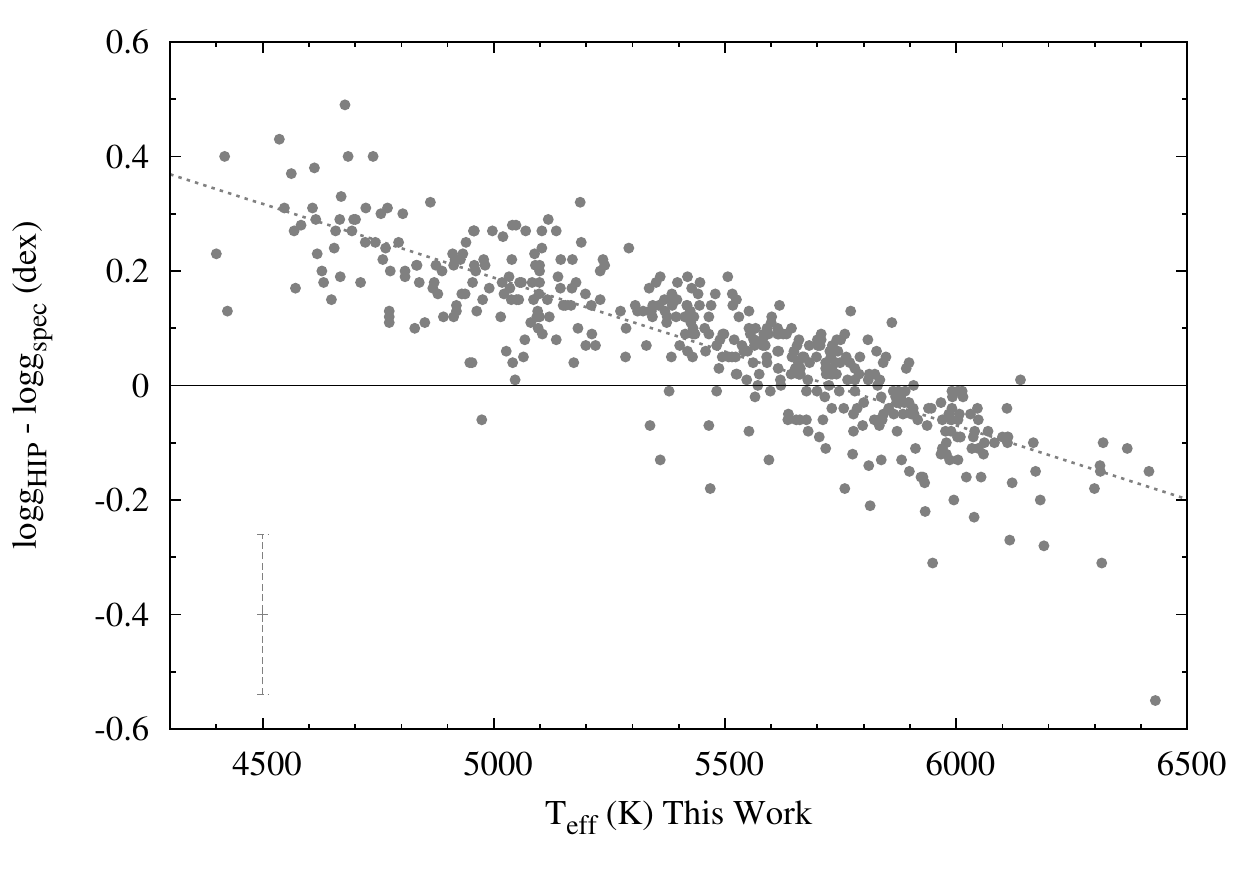}
  \caption{\footnotesize{Trigonometric $\log g_{HIP}$ minus $log g_{spec}$ as a function of temperature. The dashed line represents 
a linear fit (-2.578$\cdot$10$^{-4}$$T{}_{\mathrm{eff}}$+1.477). The mean error is shown at the bottom left.}}
  \label{logg_hip_temp}
  \end{figure}

An interesting fact is that the differences between the trigonometric and spectroscopic $\log g$ are greater for the cooler stars 
(Fig.~\ref{logg_hip_temp}). There is a clear trend between the differences in $\log g$ and $T{}_{\mathrm{eff}}$, where the underestimation 
of surface gravity in low temperatures becomes higher ($<\log g_{HIP} - \log g_{spec}>$=0.22 dex for stars with $T{}_{\mathrm{eff}}<$ 
5000\,K). This is translated into systematically higher \ion{Fe}{ii} abundances over \ion{Fe}{i} for low $T{}_{\mathrm{eff}}$. 

Such differences between \ion{Fe}{i} and \ion{Fe}{ii} abundances are difficult to explain with model uncertainties and departures from 
LTE in the spectral line formation calculations as they are expected for the warmer stars of our sample where most iron is 
ionized \citep{lind}. In addition, other model uncertainties related to granulation and activity of K-stars have been proposed to explain these 
differences even though these effects should be evident for young stars \citep{morel,schuler10}. 

Other possible explanations for these differences have to do with the iron ionization method itself. We use \ion{Fe}{ii} lines that strongly 
depend on surface gravity. For solar-type stars, however, these lines are not sufficiently present leading to poorly constrains of $\log g$. 
On the other hand, the numerous \ion{Fe}{i} lines are insensitive to $\log g$ changes. 

We have to note though, that temperatures and metallicities derived using the ionization and excitation equilibrium of iron lines are shown 
to be mostly independent of the adopted surface gravity \citep{torres}. Hence, the temperatures and metallicities derived with our 
spectroscopic method can be used as reference even if the derived spectroscopic surface gravities differ from the trigonometric values 
\citep[see also][]{santos13}. 

\section{Comparison with other methods}\label{other}

In order to evaluate the consistency of our results, namely for $T{}_{\mathrm{eff}}$, we compare them with other techniques. 
Here, we present a comparison with two different methods that are considered to be less model dependent, the Infrared Flux Method (IRFM) 
and interferometry, respectively. 

\subsection{The Infrared Flux Method - IRFM}

 \begin{figure}
  \centering
  \includegraphics[width=1.0\linewidth]{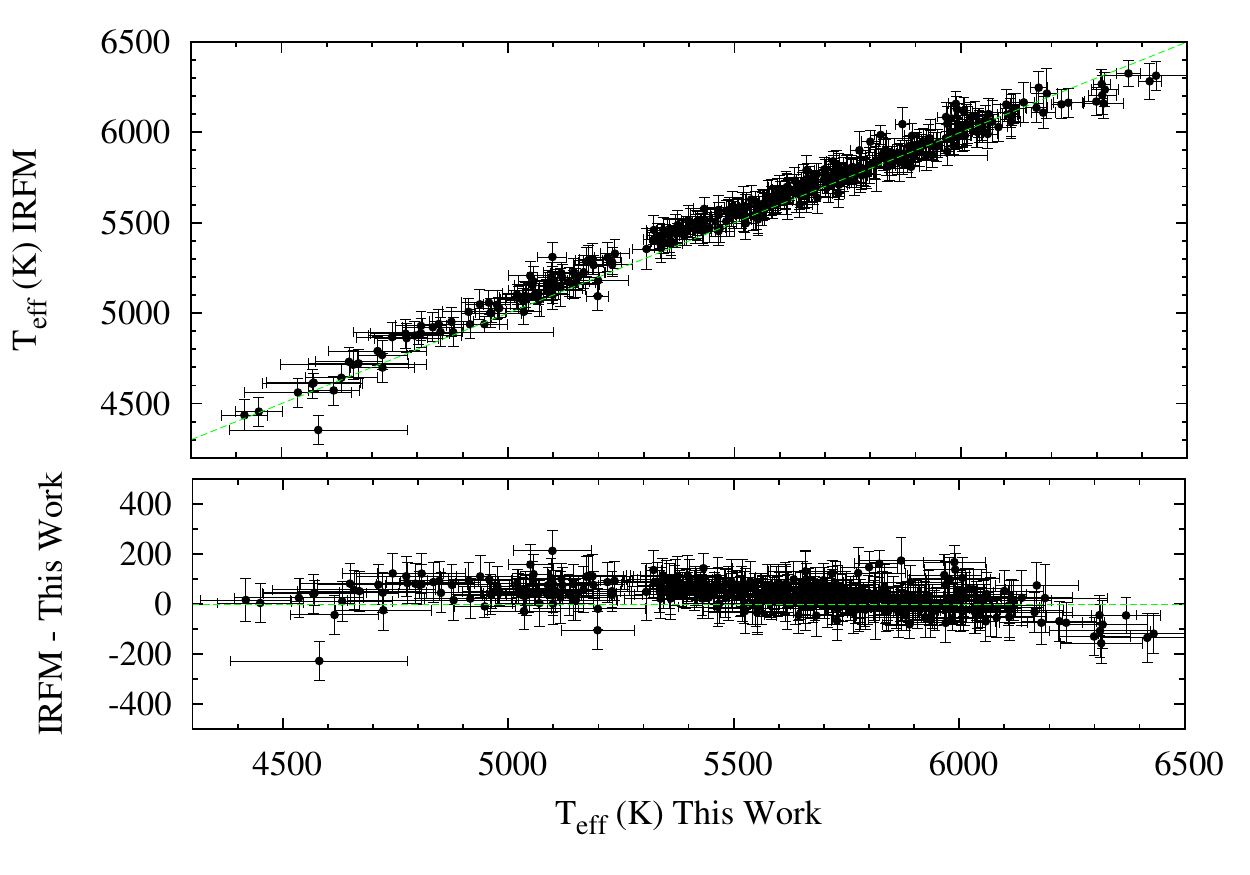}
  \caption{\footnotesize{Comparison between the temperatures derived from this work and the IRFM for stars in common.}}
  \label{fig:3}
  \end{figure}

The IRFM method \citep{black} is a semi-direct method for determining stellar parameters. The principle of this method relies on the 
fact that the bolometric flux depends on the angular diameter and the effective temperature, as described by the 
Stefan-Boltzmann law, whereas the monochromatic flux in the infrared (IR) depends on the angular diameter but weakly on the effective 
temperature, this way the dependence on the angular diameter disappears:

\begin{equation}
\centering
 \frac {f_{bol}}{f_{\lambda IR}} = \frac{\sigma T{}_{\mathrm{eff}}^{4}}{f_{\lambda IR}(model)} ,
\end{equation}

where $f_{bol}$ is the measured bolometric flux, $f_{\lambda IR}$ is the measured monochromatic IR flux and $f_{\lambda IR}$(model) is the 
monochromatic flux in the IR derived by the assuming model. The IRFM has the advantage that the dependence on the models is limited while 
the spectroscopic effective temperatures have considerable model dependence.

We compare our results with the work of \cite{casa10, casa2} that implement the IRFM method for a large sample of stars. 
The authors estimate the bolometric flux from multi-band photometric measurements in the optical \textit{BV(RC)$_{C}$} band
and in the near-IR 2MASS \textit{JHK$_{S}$} band. For the missing spectral regions, the flux is calculated by synthetic spectra computed 
from model atmospheres. The absolute calibration of Vega is based on its synthetic spectrum with an uncertainty of the zero point of 
$\sim$15 K \citep{casa10}. 

  \begin{table}
     \centering
     \caption[]{Comparison between the effective temperatures derived with different methods. $\sigma$ represents the standard deviation 
and N the number of stars for the comparison.}
     \label{total_comparison}
$$    
     \begin{array}{p{0.5\linewidth}l c c c}

           \hline\hline
               
            Method & \Delta T{}_{\mathrm{eff}} (K) & \sigma (K) & N \\
      
         \hline
IRFM -- this work & +33\pm 3 & 54 & 347 \\
IRFM -- SO08 & +14 \pm 3 & 61 & 347 \\
\hline
IRFM -- this work ($T{}_{\mathrm{eff}}<$5000\,K) & +42 \pm 12 & 65 & 29 \\ 
IRFM -- SO08 ($T{}_{\mathrm{eff}}<$5000\,K) & -86 \pm 16 & 86 & 29 \\
\hline
Interferometry -- this work & -4 \pm 33 & 98 & 9 \\
Interferometry -- SO08 & -56 \pm 35 & 105 & 9 \\
Interferometry -- IRFM & -98 \pm 83 & 204 & 6 \\
          \hline
        \end{array}
$$    
  \end{table}

Figure~\ref{fig:3} depicts the comparison between the spectroscopic temperatures and the IRFM for the stars in common. 
Temperatures of 341 stars are taken from \cite{casa2} using stars with direct application of the IRFM (\textit{irfm} sample) and 
stars with $T{}_{\mathrm{eff}}$ derived from colour calibrations (\textit{clbr} sample). Moreover, 6 stars were taken from \cite{casa10}. 
The comparison between the results of this work and the IRFM shows good agreement for all temperature ranges. In particular, for the cooler 
temperature region, the differences in $T{}_{\mathrm{eff}}$ between this work and the IRFM are much smaller and more homogeneously 
distributed than between SO08 and the IRFM. 

\begin{figure}
   \centering
   \includegraphics[width=1.0\linewidth]{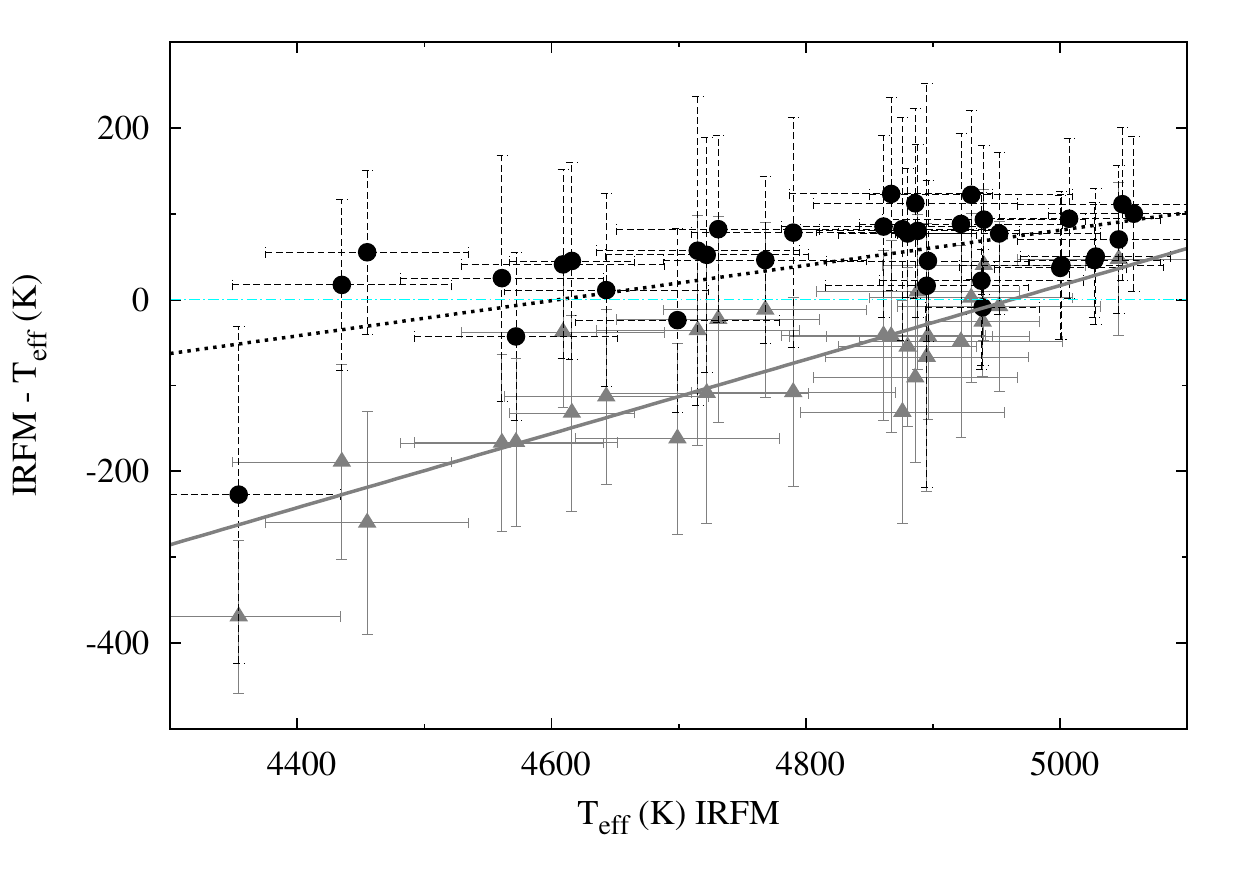}
  \caption{\footnotesize{Comparison between the difference in temperatures derived from IRFM - This Work (circles) and IRFM - SO08 
(triangles). The dashed and the solid line depict the linear fits of the data with slopes: +0.20$\pm$0.04 and +0.43$\pm$0.05, respectively.}}
  \label{fig:4}
  \end{figure}

The mean differences in temperature for the comparison samples are shown in Table~\ref{total_comparison}. It is clear that the differences 
in temperature for this work with the IRFM are constant throughout the temperature range, with a small offset of 33 K for the whole sample. 
For the cooler stars these differences are $\Delta T{}_{\mathrm{eff}}$ = 42 $\pm$ 12 K that are much smaller than SO08 with 
$\Delta T{}_{\mathrm{eff}}$ = -86 $\pm$ 16 K. Figure~\ref{fig:4} shows the comparison of stars with $T{}_{\mathrm{eff}}$$<$5000\,K. It is 
evident that the trend in $\Delta T{}_{\mathrm{eff}}$ of SO08 mostly disappears with the new temperatures. 

\subsection{Interferometry}

Precise measurements of stellar angular diameters are acquired through long baseline interferometry. The standard practice to determine the 
angular diameter is to fit the observed visibilities as a function of baseline to a uniform disk model. The angular size is connected to 
the more realistic limb darkened angular size ($\theta$$_{LD}$) using correction factors from model atmospheres \citep{claret}. Temperature 
is then derived with the standard relation: 

 \begin{equation}
 \centering
  T{}_{\mathrm{eff}} = \left( \frac{L}{4 \pi \sigma R^{2}} \right)^{1/4} =  \left( \frac{4f_{bol}}{\sigma \theta_{LD}^2} \right)^{1/4} ,
 \end{equation}
 
 where $f_{Bol}$ is usually calculated from the Spectral Energy Distribution.

\begin{table*}
\centering
\caption{Interferometric data and derived temperatures for stars in common with our sample.}
\begin{tabular}{l c c c c c c c}
\hline\hline
Star & $\theta_{LD}$ & $\Delta \theta / \theta$ & $T{}_{\mathrm{eff}}^{Int}$ & $T{}_{\mathrm{eff}}^{SO08}$ & $T{}_{\mathrm{eff}}^{this work}$ & $T{}_{\mathrm{eff}}^{IRFM}$ & References \\
 HD  & (mas) & (\%) & (K) & (K) & (K) & (K) &   \\
\hline
10700 & 2.022 $\pm$ 0.011 & 0.54 & 5383 $\pm$ 47 & 5310 $\pm$ 17 & 5322 $\pm$ 17 & 5459 $\pm$ 80 &  1, a \\
... & 1.971 $\pm$ 0.050 & 2.54 & 5449 $\pm$ 83 & ... & ... & ... & 2, a \\
11964 & 0.611 $\pm$ 0.081 & 13.25 & 5413 $\pm$ 359 & 5332 $\pm$ 22 & 5285 $\pm$ 21 & - & 3, b \\
19994 & 0.788 $\pm$ 0.026 & 3.30 & 6109 $\pm$ 111 & 6289 $\pm$ 46 & 6315 $\pm$ 44 & 6159 $\pm$ 80 & 3, b \\
22049 & 2.148 $\pm$ 0.029 & 1.35 & 5107 $\pm$ 21 & 5153 $\pm$ 42 & 5049 $\pm$ 48 & 5207 $\pm$ 80 & 4, c \\
23249 & 2.394 $\pm$ 0.029 & 1.21 & 4986 $\pm$ 57 & 5150 $\pm$ 51 & 5027 $\pm$ 48 & - & 5, a \\
26965 & 1.504 $\pm$ 0.006 & 0.40 & 5143 $\pm$ 14 & 5153 $\pm$ 38 & 5098 $\pm$ 32 & 5311 $\pm$ 80 & 6, d \\
... & 1.650 $\pm$ 0.060 & 3.63 & 4910 $\pm$ 90 & ... & ... & ... & 4, d \\
128621 & 6.001 $\pm$ 0.021 & 0.35 & 5182 $\pm$ 24 & 5234 $\pm$ 63 & 5168 $\pm$ 75 & - & 7, e \\
146233 & 0.676 $\pm$ 0.006 & 0.89& 5836 $\pm$ 46 & 5818 $\pm$ 13 & 5810 $\pm$ 12 & 5826 $\pm$ 80 & 8, f \\
... & 0.780 $\pm$ 0.017 & 2.18 & 5433 $\pm$ 69 & ... & ... & ... & 9, f \\
209100 & 1.890 $\pm$ 0.020 & 1.06 & 4527 $\pm$ 29 & 4754 $\pm$ 89 & 4649 $\pm$ 73 & 4731 $\pm$ 80 & 4, g \\
\hline
\end{tabular}
\tablefoot{References for $\theta_{LD}$: (1)~\cite{teixeira}; (2) \cite{pij};
(3) \cite{van1}; (4) \cite{kervella08}; (5) \cite{thevenin};
(6) \cite{boy2}; (7) \cite{kervella03}; (8) \cite{bazot}; (9) \cite{boy1}. 
References for $f_{bol}$: (a)~\cite{bru1}; (b) \cite{van1}; (c) \cite{cayrel}; (d) \cite{boy2};  
(e) \cite{ramirez}; (f) \cite{boy1}; (g) \cite{ramirez05}.} 
\label{interferometry}
\end{table*}

\begin{figure}
\centering
\makebox[\linewidth]{
\includegraphics[width=1.0\linewidth]{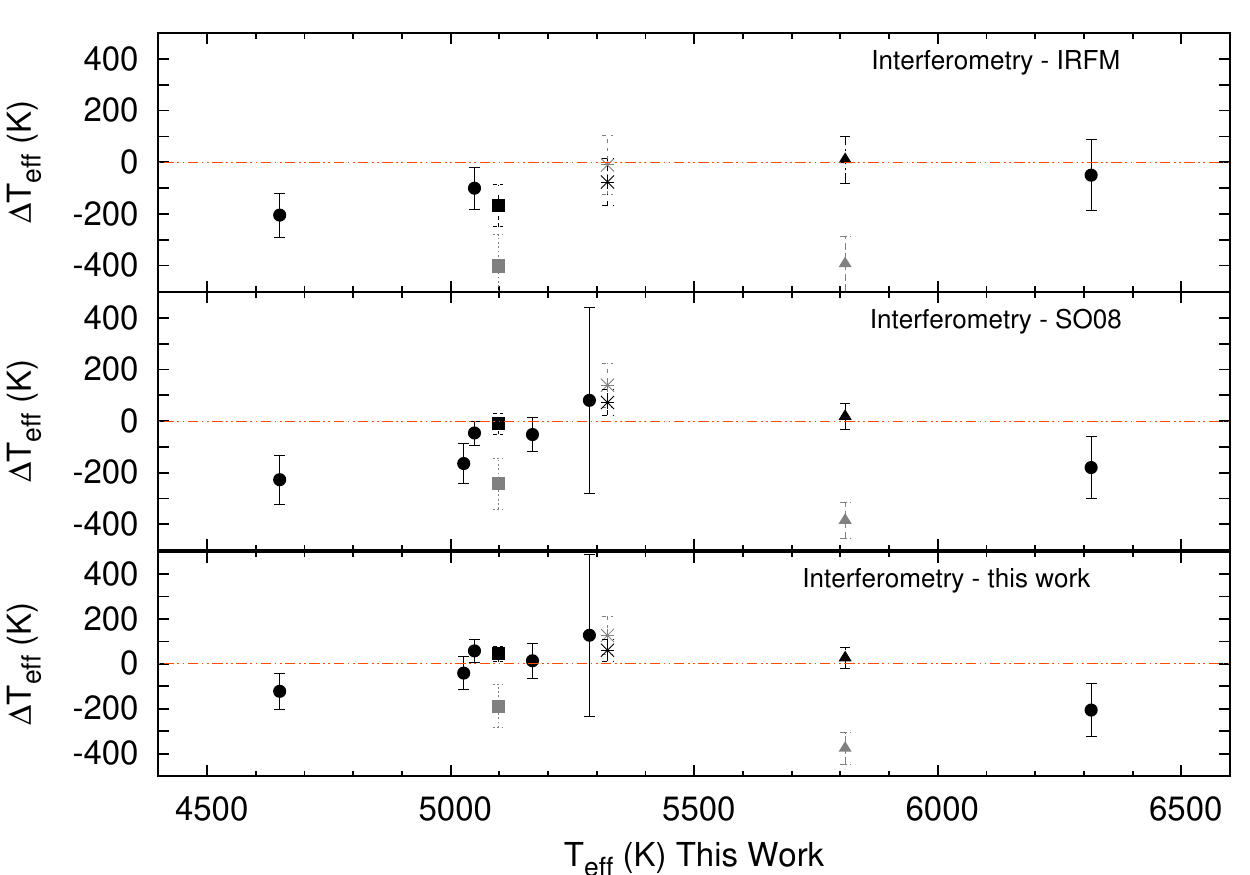}}
\caption{\footnotesize{Comparison between the spectroscopic, IRFM and the direct temperature measurements for stars in common with the 
literature. The x-axis corresponds to temperatures derived from this work. Square symbols are for the different temperatures for HD\,26965, 
stars for HD\,10700 and triangles for HD\,146233. Stars with grey colour represent $T{}_{\mathrm{eff}}$ with high angular diameter 
uncertainty.}}
\label{fig:6}
\end{figure}

We compare our results with the temperatures derived from interferometry. Unfortunately, the number of stars with available angular 
diameters for this sample is only down to a few since these measurements are challenging for dwarfs due to their small photospheric disks 
that are difficult to resolve. We have 9 stars in common for the comparison with spectroscopy and 6 with the IRFM. We use only direct 
angular diameters and bolometric fluxes available in the literature from Table~\ref{interferometry}. Our results show better agreement 
for these stars to interferometry than when comparing with the values of SO08 and the IRFM (see Table~\ref{total_comparison}). We have to 
note though, that the comparison sample is very small and the values of $T{}_{\mathrm{eff}}$ were derived from the \textit{clbr} sample 
that is not the best representative for the IRFM precision.  

The differences in temperatures for the different methods are plotted in Fig.~\ref{fig:6}. For 3 stars (HD\,10700, HD\,26965, HD\,146233) 
we include angular diameter measurements from different authors that give different $T{}_{\mathrm{eff}}$ and are represented with 
different symbols. In the same figure, we see that from the stars with multiple measurements, the ones with the smallest uncertainty 
($\frac{\Delta \theta}{\theta}$\%) agree better with the temperatures derived with the spectroscopic and photometric methods. 
A precision better than 2\% in the angular diameters corresponds to an accuracy of 1\% in the effective temperatures, which is roughly 60 K 
at solar temperature, assuming no error in the bolometric flux. It is useful thus, to take the uncertainty of the angular diameter into 
consideration for a reliable determination of temperature. 

\section{New atmospheric parameters for cool planet hosts}\label{planets}

In general, for planet host stars with effective temperature below 5200\,K, the new set of parameters, derived using the line list
presented in this paper, imply that the planet host stars have lower temperatures than previously published. This has implications
for both their mass and radius determination. The lower temperatures imply lower stellar masses as well as lower stellar radii.
This means that the derived planetary masses and radii (for transit planet cases) are also lower. As a consequence of the mass reduction,
the semi-major axis of the orbits will also be smaller. The expected effects are however small, and no major revisions are expected to
occur. 

With this new line list, we re-derive the stellar parameters for 10 ``cool'' planet hosts already published in the literature and are not 
included in the 451 stellar sample of this work. We only consider GK dwarfs with an effective temperature lower than 
5200\,K whose planets were detected with the radial velocity technique from the CORALIE and HARPS GTO planet search samples. These planet 
hosts have been previously analyzed with high S/N spectra following the same procedure as this work but with different line lists. 

In Table \ref{TabHosts}, the fundamental parameters based on the new line list are presented. The sixth column gives the reference of the 
previously published parameters. 
To explain the effect of these new parameters to mass more quantitative, we calculate the stellar masses from the Padova interface for 
both with the original and new parameters. We avoid to use the published stellar masses in order to compare uniformly. We find the maximum 
difference in mass to be 1.5\% in absolute units which is negligible compared to the standard mass error.

\begin{table*}
\caption{Updated stellar parameters for previously analyzed planet hosts.}
\label{TabHosts}
\centering
\begin{tabular}{lccccc}
\hline\hline
Name & $T{}_{\mathrm{eff}}$ & $\log g$ & [Fe/H] & $\xi_{t}$ & Reference \\
 & (K) & (cm s$^{-2}$) & (dex) & (km s$^{-1}$) & \\
\hline
BD-082823 & 4648 $\pm$ 135 & 4.33 $\pm$ 0.32 & 0.00 $\pm$ 0.08 & 0.27 $\pm$ 0.81 & 1 \\
HD 3651 & 5182 $\pm$ 79 & 4.30 $\pm$ 0.16 & 0.12 $\pm$ 0.05 & 0.66 $\pm$ 0.15 & 2 \\
HD 13445 & 5114 $\pm$ 61 & 4.55 $\pm$ 0.13 & -0.29 $\pm$ 0.04 & 0.66 $\pm$ 0.15 & 2 \\
HD 20868 & 4720 $\pm$ 91 & 4.24 $\pm$ 0.22 & 0.08 $\pm$ 0.06 & 0.47 $\pm$ 0.31 & 1 \\
HD 99492 & 4815 $\pm$ 184 & 4.28 $\pm$ 0.46 & 0.24 $\pm$ 0.12 & 0.50 $\pm$ 0.56 & 4 \\
HD 125595 & 4596 $\pm$ 235 & 4.25 $\pm$ 0.63 & 0.10 $\pm$ 0.14 & 0.14 $\pm$ 1.41 & 3 \\
HD 128311 & 4778 $\pm$ 75 & 4.35 $\pm$ 0.17 & -0.03 $\pm$ 0.02 & 0.82 $\pm$ 0.16 & 2 \\
HD 192263 & 4906 $\pm$ 57 & 4.36 $\pm$ 0.17 & -0.07 $\pm$ 0.02 & 0.78 $\pm$ 0.12 & 2 \\
HD 215497 & 5003 $\pm$ 103 & 4.26 $\pm$ 0.26 & 0.25 $\pm$ 0.05 & 0.61 $\pm$ 0.22 & 1 \\
HIP 5158 & 4673 $\pm$ 175 & 4.24 $\pm$ 0.47 & 0.22 $\pm$ 0.12 & 0.34 $\pm$ 1.09 & 1 \\
\hline
\end{tabular}
\tablebib{
(1) ̃\cite{sousa3}; (2) \cite{san1}; (3) \cite{Seg11}; (4) \cite{San05}.
}

\end{table*}

\section{[\ion{Cr}{i}/\ion{Cr}{ii}] and [\ion{Ti}{i}/\ion{Ti}{ii}] vs. $T{}_{\mathrm{eff}}$ with the new atmospheric parameters}\label{elements}

The precise and accurate stellar parameters are, as well, very important for further analyzing stellar chemical abundances. The traditional 
spectroscopic abundance analysis methods require these parameters as input to compute the atmosphere models, hence the accuracy 
of the final elemental abundances depends on the accuracy of these input parameters. Different atoms and ions are not equally sensitive to 
all the stellar parameters. For example, ionized species are more sensitive to gravity variations than neutral species 
\citep[e.g.][]{gilli, neves, adib}.

Recently, \cite{neves} and \cite{adib} analyzing chemical abundances of the refractory elements of the HARPS sample stars,
observed some unexpected trends with effective temperature. Particularly, they detected systematic trends of [X/H] or [X/Fe] with 
$T{}_{\mathrm{eff}}$ for some elements at low temperatures and found that [\ion{Cr}{i}/\ion{Cr}{ii}] and [\ion{Ti}{i}/\ion{Ti}{ii}]
abundance ratios gradually increase with decreasing effective temperature when $T{}_{\mathrm{eff}} \lesssim 5000$\,K.
Similar trends for different elements with \emph{$T{}_{\mathrm{eff}}$} have been already noted in the literature 
\citep[see e.g.][]{val, preston, gilli, lai, suda}. Different explanations of the mentioned trends are discussed 
in the literature. The unexpected trends in the low temperature regime may be due to the stronger line blending and may also be 
connected to either deviations from excitation or ionization equilibrium, or to problems associated with the differential analysis 
\citep{neves}. A possible explanation for the observed trends with \emph{$T{}_{\mathrm{eff}}$} could also be an incorrect 
T-$\tau$ relationship in the adopted model atmospheres \citep{lai} or NLTE effects \citep{bodaghee}. Summarizing, it can be assumed that 
the observed trends are probably not an effect of stellar evolution, and uncertainties in atmospheric models are the dominant effect 
in measurements (see also the discussion in \citealt{adib}).

\begin{figure}
\centering
\includegraphics[width=1.0\linewidth]{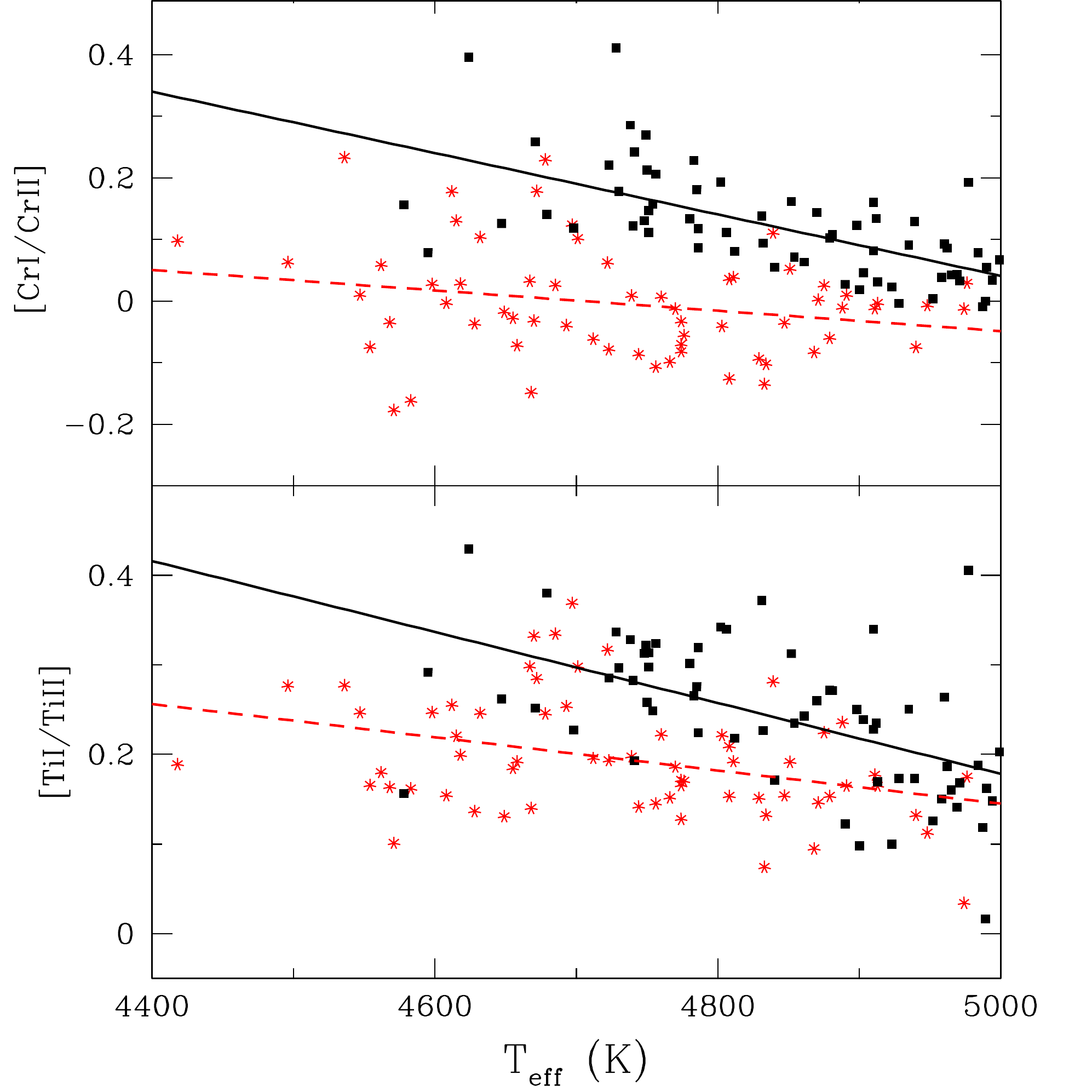}
\caption{[\ion{Cr}{i}/\ion{Cr}{ii}] and [\ion{Ti}{i}/\ion{Ti}{ii}] as a function of effective temperature. The black squares and red asterisks
correspond to the abundance ratios derived using old and new stellar parameters, respectively. The solid and dashed lines depict the linear 
fits of the data.}
\label{fig:7}
\end{figure}

In Fig.~\ref{fig:7}, we plot the [\ion{Cr}{i}/\ion{Cr}{ii}] and [\ion{Ti}{i}/\ion{Ti}{ii}] abundance ratios derived using the stellar 
parameters of SO08 and this work as a function of the $T{}_{\mathrm{eff}}$ for stars cooler than 5000\,K. This plot is useful to ensure 
that the ionization equilibrium enforced on the \ion{Fe}{ii} lines is acceptable to other elements. As can be seen the slopes of 
the abundance ratios with new parameters are very gentle. The new slope of [\ion{Cr}{i}/\ion{Cr}{ii}] per 1000 K is $-0.16\pm0.08$,
whereas the slope with the parameters of SO08 is $-0.49\pm0.08$. The new slope of [\ion{Ti}{i}/\ion{Ti}{ii}] is also improved a lot and 
is $-0.18\pm0.06$ dex per 1000 K. For comparison the slope of SO08 is $-0.39\pm0.08$ dex. Although the trend with $T{}_{\mathrm{eff}}$ is
weak, there is a shift of about 0.2 dex for the [\ion{Ti}{i}/\ion{Ti}{ii}] ratio. This shift is difficult to connect to the still possible 
uncertainties in the stellar parameters and its exact nature still remains to be clarified. Probably, one (or more) of the above mentioned 
effects can be responsible for that. Unfortunately, in the literature there is no available NLTE calculations for the \ion{Ti}{i} and 
\ion{Ti}{ii} lines used in our study, and it is difficult to estimate the NLTE effect for the [\ion{Ti}{i}/\ion{Ti}{ii}] ratio. 

Summarizing, this independent test shows that the new stellar parameters derived from the iron lines more carefully chosen 
for cooler stars make the observed [\ion{Cr}{i}/\ion{Cr}{ii}] and [\ion{Ti}{i}/\ion{Ti}{ii}] trends with $T{}_{\mathrm{eff}}$ much weaker. 

\section{Conclusions}\label{conclusion}

In this work, we present a new iron line list in order to correct for discrepancies in temperatures for a sample of 451 stars that is 
part of the HARPS GTO program using the cool star HD\,21749 as a reference. The quality of the line list plays a key role for temperature 
determination especially for the K type stars. The new line list is compiled in order to eliminate blended lines in the spectra of 
these stars that suffer more from such effects. We also apply a limitation to very strong and very weak lines that usually cause errors in 
the EW measurements (see Sect.~\ref{build}).

We derived the stellar parameters for the 451 stars of the sample in a homogeneous way with the new line list and compare our results with 
the work of SO08, where the authors followed the same analysis but with an expanded line list.  

We find very good agreement for the high and intermediate ranges in temperature with SO08. The differences appear in the lower temperatures 
below $\sim$5000\,K. In addition, surface gravity and metallicitity remain unaffected with the new line list. These results are very 
important since accurate parameters of planet host stars are essential to characterize the planets. In particular, the metallicities of 
this sample are used for the investigation between the stars and their planet hosts. The fact that the new metallicities are not different 
from the old ones supports the already established correlation between the planet host stars and planet frequency. 

The agreement with the IRFM that is considered to be a less model dependent technique, suggests that our parameters are more precise at 
least for the cooler stars compared to the previous study of SO08 where the existing trend in low temperatures between the 
spectroscopic and the photometric temperatures disappears with the new line list. In addition, we compare our new temperatures with the 
ones derived with interferometry for 9 star with available angular diameter measurements. Even though the sample is very small, a 
comparison between these results shows very good agreement. Attention should be made to temperatures derived with the angular diameters 
that have larger uncertainties. 

Having checked the reliability of the new line list, we apply it to determine fundamental parameters for some of the planet 
hosts with low temperatures, which were previously analyzed with different line lists with available high S/N spectra. Finally, we use 
the new parameters for stars with $T{}_{\mathrm{eff}}<$5000\,K to significant reduce the trends of [\ion{Cr}{i}/\ion{Cr}{ii}] and 
[\ion{Ti}{i}/\ion{Ti}{ii}] abundance ratios with temperature.

\begin{acknowledgements}
The authors acknowledge the referee, I. Ram\'{i}rez, for his comments that helped improve this paper. 
This work was supported by the European Research 
Council/European Community under the FP7 through Starting Grant agreement number 239953. 
N.C.S. also acknowledges the support from Funda\c{c}\~ao para a Ci\^encia e a Tecnologia (FCT) through program Ci\^encia 2007 funded by FCT/MCTES 
(Portugal) and POPH/FSE (EC), and in the form of grant reference PTDC/CTE-AST/098528/2008. V.Zh.A. and S.G.S. are supported by grants 
SFRH/BPD/70574/2010 and SFRH/BPD/47611/2008, respectively. 
\end{acknowledgements}

\bibliography{bibliography}

\begin{appendix} 

\section{Reduced EW versus Excitation Potential}\label{app_re}

\begin{figure}
\centering
\includegraphics[width=1.0\linewidth]{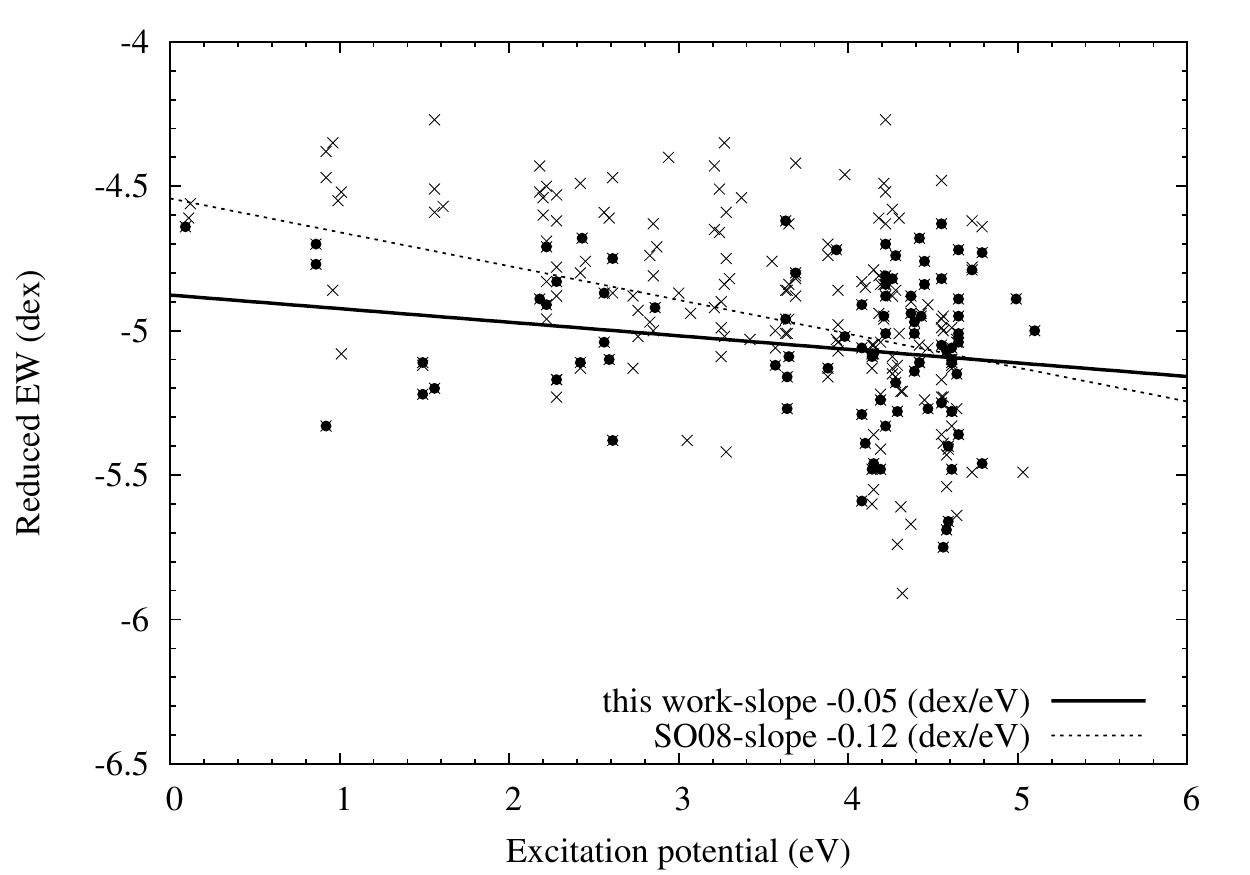}
\caption{Reduced EW versus excitation potential for the line list of this work (circles) and for SO08 (crosses). The slope of 
the linear fits is also depicted.}
\label{ew_ep}
\end{figure}

We show the correlation between the reduced EW and excitation potential (Fig.~\ref{ew_ep}). For the line list of this work there is 
much smaller correlation than in the work of SO08 suggesting small degeneracies between $\xi_{t}$ and $T{}_{\mathrm{eff}}$.

\end{appendix}

\begin{appendix} 

\section{The microturbulence relationship}\label{app}

Microturbulence is taken into consideration for abundance analyses to reconcile differences between the observed and predicted from models 
equivalent widths of strong lines. Previous studies of FGK dwarfs have shown that $\xi_{t}$ depends on $T{}_{\mathrm{eff}}$ and 
$\log g$ \citep[e.g.][]{nissen,reddy,allende, adib2, ramirez13}. 
Using a linear regression analysis to the new parameters of the sample, we derive the following expression:

\begin{equation}\label{mic} 
 \xi_{t} = 6.932 \times 10^{-4} T{}_{\mathrm{eff}} - 0.348 \log g - 1.437 . 
\end{equation}

Here, $\xi_{t}$ is in km/s, $T{}_{\mathrm{eff}}$ and $\log g$ are in their traditional units. The parameters of the stars in the 
sample range: 

4400 $< T{}_{\mathrm{eff}} <$ 6400 K,

3.6 $< \log g <$ 4.8 dex, 

-0.8 $<$ [Fe/H] $<$ 0.4 dex. 

\begin{figure}
\centering
\includegraphics[width=1.0\linewidth]{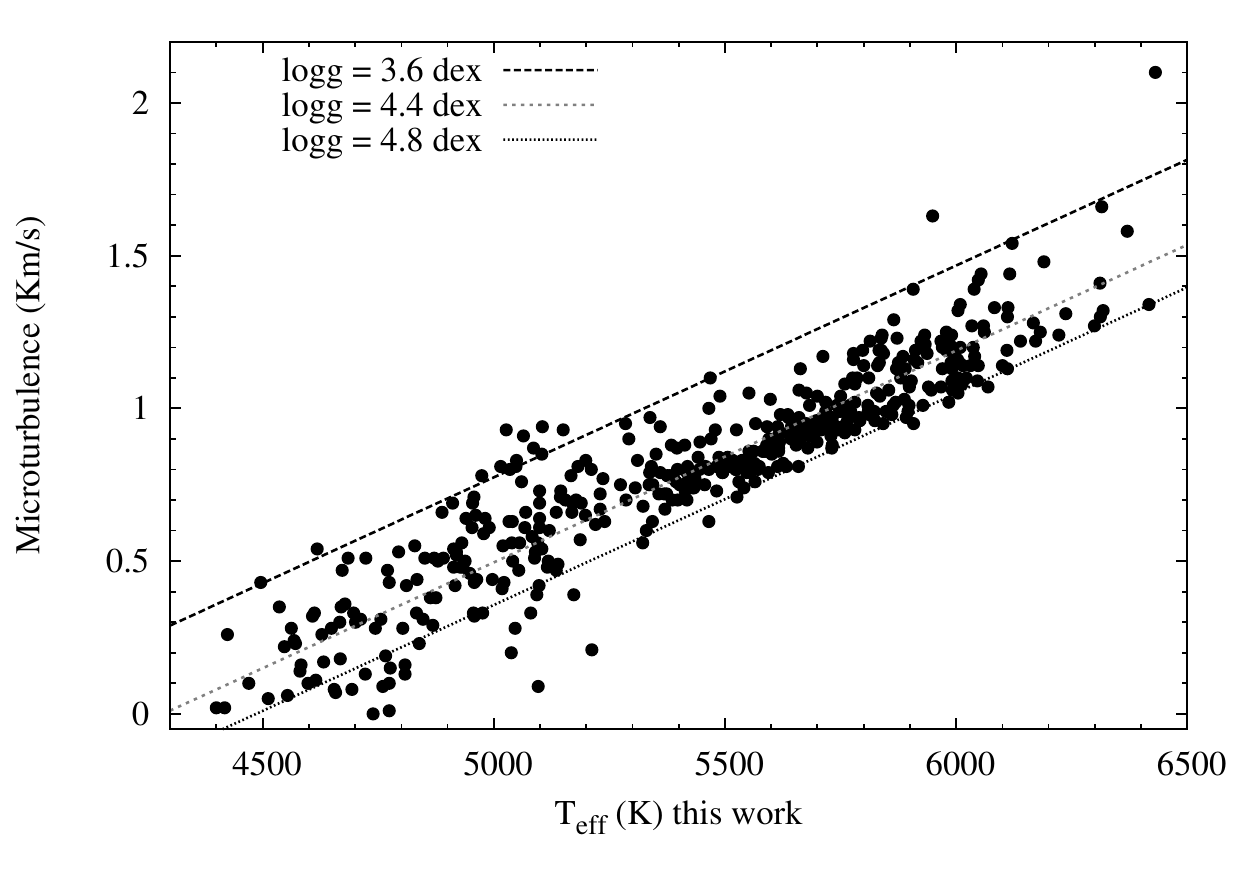}
\caption{Correlation of microturbulence with temperature and surface gravity as described by equation~\ref{mic}. We set $\log g$ to the 
highest, average and lowest values of the sample.}
\label{micro}
\end{figure}

The new derived parameters indicate a linear dependence on temperature for a set value of surface gravity.  
In Fig.~\ref{micro}, we see the dependence of microturbulence on temperature for a set of $\log g$ values. Microturbulence clearly 
increases with temperature and decreases with surface gravity.

\end{appendix}

\end{document}